\documentclass[12pt]{article}
\usepackage{graphicx}
\begin{document}
\rightline{NKU-04-SF1}
\bigskip
\begin{center}
{\Large\bf Greybody Factors  of  Charged  Dilaton Black Holes 
 in 2 + 1 Dimensions 
}

\end{center}
\hspace{0.4cm}
\begin{center}
Sharmanthie Fernando \footnote{fernando@nku.edu}\\
{\small\it Department of Physics \& Geology}\\
{\small\it Northern Kentucky University}\\
{\small\it Highland Heights}\\
{\small\it Kentucky 41099}\\
{\small\it U.S.A.}\\

\end{center}

\begin{center}
{\bf Abstract}
\end{center}

\hspace{0.7cm} 

We have studied  scalar perturbations of charged dilaton black holes in 2+1 dimensions. The black hole considered here is a solution to the low-energy string theory in 2+1 dimensions. The exact decay rates and the grey body factors for the massless minimally coupled scalar is computed for both the charged and the uncharged dilaton black holes. The charged and the uncharged black hole show  similar behavior  for grey body factors, reflection coefficients and decay rates. 


\section{Introduction}

According to the seminal work of Hawking \cite{haw1} \cite{haw2}, black holes emits radiation from its horizon. However, it is not a perfectly thermal black body radiation. The gravitational potential surrounding the horizon scatter some of the radiation back into the black hole and transmit some to infinity leading to a frequency dependent greybody factor. Some of the earlier work  on absorption cross sections are given in \cite{star} \cite{gib} \cite{page} \cite{unru}.

In recent times the greybody factors of black holes have attracted attention due to various reasons, one of them  being the string computations. There has been considerable progress in describing the properties of black holes from string theory. For example, the Bekenstien-Hawking black hole entropy formula was derived for certain class of black holes by counting D-brane bound states \cite{vafa}. In an attempt to understand it further, much work has been done to compare the details of string computation with that of the classical computations of black holes. 
There have been several work to show that the greybody factors and decay rates obtained using D-branes agree exactly with the classical calculations of scalar perturbations of black holes. Some of the  work in this context are \cite{mal1} \cite{dhar} \cite{das1} \cite{gubser1} \cite{mal2} \cite{gubser2} \cite{gubser3} \cite{kleb} \cite{alberto}.

Another interesting aspect of the absorption cross section was given by \cite{das2}. There, the low energy absorption cross section for a massless scalar was shown to be proportional to the area of the horizon. This Universality was shown to be maintained for higher dimensional objects such as black p-branes \cite{emparan} and for rotating black holes \cite{mal3} \cite{lar}. A study along these lines on stationary, circularly symmetric black holes were done by Higuchi \cite{hig}.

In an extension of this work, Jung et.al.\cite{jung} studied the absorption cross section for massive scalars and showed that the Universality is valid only under certain restrictions of the parameters involved.

In this paper, we study the grey body factors and decay rates of a charged dilaton black hole in three dimensions. The well known BTZ black hole \cite{banados} has been studied extensively on the topics described above. Birmingham et. al.  \cite{bir1} studied the BTZ black hole grey body factor for massless scalars and proved the agreement with the low-energy Universal law in \cite{das2}. The greybody factors of the BTZ black hole for various scalars were studied in \cite{lee1}. The greybody factors of BTZ has been studied from the point of 5D black holes since 5D black hole is U-dual to $BTZ \times S^3$ \cite{das3} \cite{lee2}. Another interesting work in three dimensions were to study absorption cross section of the black string \cite{lee3} and to study the massive scalar absorption in de-Sitter space \cite{myong}.

Here, we study how the greybody factors are effected when charge and a dilaton is introduced for the BTZ black hole. We have chosen the static charged black hole derived by Chan and Mann \cite{chan}. This black hole is appealing due to the fact that it arises from the low energy string theory in three dimensions with a negative cosmological constant.

We have organized the paper as follows: In section 2 an introduction to the geometry of the black hole is given. The scalar perturbation of the black hole is given  in section 3. The grey body factors and decay rates are considered for the uncharged black hole in section 4 and for the charged black hole in section 5. Finally the conclusion is given in section 6.

\section{ Geometry of the static charged dilaton black hole}

In this section we will present the geometry and important details of the staic charged black hole. The Einstein-Maxwell-dilaton action which lead to these black holes considered by Chan and Mann \cite{chan} is given as follows:
\begin{equation}
S = \int d^3x \sqrt{-g} \left[ R - 4  (\bigtriangledown \phi )^2 -
e^{-4  \phi} F_{\mu \nu} F^{\mu \nu} + 2 e^{4 \phi} \Lambda \right]
\end{equation}
Here, $ \Lambda$ is treated as the cosmological constant. ( $\Lambda > 0$ anti-de Sitter and  $\Lambda < 0$ de Sitter). $\phi$ is the dilaton field, $R$ is the scalar curvature and $F_{\mu \nu}$ is the Maxwell's field strength. This action is conformally related to the low-energy string action in 2+1 dimensions. The static circularly symmetric solution to the above action is given by,

$$
ds^2= - f(r)dt^2 +  \frac{4 r^2 dr^2}{f(r)} + r^2 d \theta^2
$$
\begin{equation}
f(r) =\left( -2Mr + 8 \Lambda r^2 + 8 Q^2 \right); \hspace{0.1cm} \phi = \frac{1}{4}  ln (\frac{r}{\beta}) ; \hspace{1.0cm}F_{rt} = \frac{Q}{r^2}
\end{equation}
For $M \geq 8 Q \sqrt{\Lambda}$, the space-time represent a black hole. It has two horizons given by the zeros of $g_{tt}$;
\begin{equation}
r_+ =  \frac{M + \sqrt{ M^2 - 64 Q^2 \Lambda}}{8 \Lambda}; \hspace{1.0cm}
r_- = \frac{M - \sqrt{ M^2 - 64 Q^2 \Lambda}}{8 \Lambda}
\end{equation}
There is a singularity at $r=0$ and it is time-like. An important thermodynamical quantity corresponding to a black hole is the  Hawking temperature $T_H$. It is given by,
\begin{equation}
T_H= \frac{1}{4 \pi} |\frac{dg_{tt}}{dr}| \sqrt{-g^{tt} g^{rr}} |_{r=r_+} = \frac{M}{4 \pi r_+} \sqrt{ 1 - \frac{64 Q^2 \Lambda}{M^2}}
\end{equation}
The temperature $T_H=0$ for the extreme black hole with $M= 8 Q \sqrt{\Lambda}$. For the uncharged black hole $T_H = \frac{\Lambda}{ \pi}$.

This black hole is also a solution to low energy string action  by a  conformal transformation,
\begin{equation}
g^{String} =  e^{4 \phi} g^{Einstein}
\end{equation}
In string theory, different space-time geometries can be related to each other by duality transformations. 
The charged black hole in eq.(2) is dual to an uncharged ``black string'' given by the following space-time geometry,
$$
ds^2_{Einstein} = - ( 8 \Lambda \beta r - 2 m \sqrt{r})  dt^2 +  \frac{dr^2}{( 8 \Lambda \beta r - 2 m \sqrt{r})  } + \gamma^2 r d \theta^2
$$
\begin{equation}
\Phi = -\frac{1}{4} ln ( \frac{r}{\beta})
\end{equation}
Here $m^2$ is the mass per unit length and $\gamma$ is an integration constant. It is an uncharged solution of the action in eq.(1). Before discussing the origin of this particular duality, let us mention the history of the  uncharged black hole in eq.(6). Mandal et.al. \cite{mandal} found a (1+1) dimensional black hole in string theory given below:  
$$
ds^2_{Einstein} = - ( 1  - \frac{M}{r})  dt^2 +  \frac{k dr^2}{ 4r ( r -M)  } 
$$
\begin{equation}
\Phi = -\frac{1}{2} ln r - \frac{1}{4} ln k
\end{equation}
Here, $M$ is the mass of the black hole and $k$ is a constant.  This (1+1) (MSW) black hole is widely studied. For example, Witten has found the exact conformal field theory corresponding to it \cite{witten}. By taking the product of (1+1) MSW black hole with $\bf{R}$, yields an uncharged ``black string'' solution given in eq.(6). 

In string theory,  is possible to create charged solutions from uncharged ones in by duality transformations. For a review of such transformation see Horowitz\cite{horo1}. It is possible to apply the following  transformation to a given  metric in string frame to obtain a charged version.
$$ \hat{g}_{tt} = \frac{ g_{tt}}{[ 1 + (1 + g_{tt}) Sinh^2 \alpha ]^2}$$
$$\hat{A}_t = - \frac{ ( 1 + g_{tt}) Sinh 2\alpha )}{2 \sqrt{2} [ 1 + (1 + g_{tt}) Sinh^2 \alpha ]}$$
\begin{equation}
e^{- 2 \hat{\phi}} = e^{ -2 \phi} [ 1 + (1 + g_{tt}) Sinh^2 \alpha ]
\end{equation}
Note that $g_{tt}$ is in the string frame. Here $\alpha$ is an arbitrary parameter. Now, by taking the metric in eq.(6) in string frame and performing the above transformations yields a dual  metric as follows:

$$
ds^2_{Einstein} = - ( 8 \Lambda \beta r - 2 m \sqrt{r})  dt^2 +  \frac{ P(r)dr^2 }{( 8 \Lambda \beta r - 2 m \sqrt{r})  } + \gamma^2 r  P(r)d \theta^2$$
\begin{equation}
e^{-2\phi} = \sqrt{ \frac{r}{\beta} } P(r)
\end{equation}
Here,
\begin{equation}
P(r) =  1 + (1 - 8 \Lambda \beta^2) Sinh^2 \alpha  + \frac{ 2 m \beta Sinh^2 \alpha}{\sqrt{r}}
\end{equation}
Note that the dual metric is given in the Einstein frame. By performing a coordinate transformation given by $\hat{r} = P(r)^2 r$ and replacing $\hat{r}$ with $r$ yields the charged black hole in eq.(2). Hence the static charged black hole considered in this paper is dual to the uncharged (2+1) MSW black hole. Note that the  transformation given in eq.(8) is a part of $O(2,1)$ symmetry group of the low energy sting action which is described well in Sen et.al.\cite{sen}.  There is another transformation which is a part of $O(2,1)$ group known as ``space-time'' or ``target space'' duality. Performing such transformation on the rotating BTZ black hole,  Horowitz and Welch \cite{horo2} obtained another charged black string solution in 2+1 dimensions given as follow;
$$
ds^2_{Einstein} = - \left( 1 -  \frac{M} {r}\right)  dt^2 +  \left( 1 -  \frac{Q^2} {Mr} \right)^{-1}  dx^2 + \left( 1 -  \frac{M} {r} \right) \left( 1 -  \frac{Q^2} {Mr} \right)^{-1}  \frac{ l^2 dr^2 }{4 r^2  } $$
\begin{equation}
e^{-2\phi} =  ln rl, \hspace{1.0cm} B_{xt} = \frac{Q}{r}
\end{equation}
Since the transformations used to obtain the two charged black holes are different, the two have distinct properties. How ever, when $Q=0$ the above black hole corresponds to the uncharged black hole in eq.(6).  Therefore, by applying  two distinct elements from the $O(2,1)$ symmetry group to the static BTZ black hole,
the charged black string considered in this paper can be constructed.

%%%%%%%%%%%%%%%%%%%%%%%%%%%%%%%%%

\section{Scalar perturbation of charged black holes}

In this section we will develop the equations for a scalar field in the background of the static charged black hole introduced in the previous section. The general equation for a massless scalar field in curved space-time can be written as,
\begin{equation}
\bigtriangledown ^2 \Phi  =0
\end{equation}
which is also equal to,
\begin{equation}
\frac{1}{\sqrt{-g}} \partial_{\mu} ( \sqrt{-g} g^{\mu \nu} \partial_{\nu} \Phi )=0 
\end{equation}
Using the anzatz,
\begin{equation}
\Phi =  e^{- i \omega t} e^{i m \theta} R(r) 
\end{equation}
in eq.(13) leads to the radial equation,
\begin{equation}
\frac{d}{dr} \left( \frac{f(r)}{2} \frac{dR(r)}{dr} \right) + 2r^2 \left( \frac{\omega^2}{f(r)}   - \frac{m^2}{r^2}  \right)  R(r) =0
\end{equation}
If we redefine $R(r)$ as
\begin{equation}
R(r) = \frac{ \xi(r)}{\sqrt{r}}
\end{equation}
and use the tortoise coordinate $r_{*}$ given by 
\begin{equation}
dr_{*} = \frac{2 r dr}{f(r)}
\end{equation}
the above wave equation simplifies to a one dimensional Schrodinger equation,
\begin{equation}
\left(\frac{d^2 }{dr_*^2} + \omega^2 - V(r) \right) \xi(r_*) = 0
\end{equation}
Here, $V(r)$ is given by,
$$
V(r) = \frac{ m^2 f}{r^2} + \frac{d}{dr} \left( \frac{ f}{ 4 r^{3/2}} \right) \frac{ f}{ 2 r^{3/2}}
$$
\begin{equation}
= -\frac{12 Q^4}{r^2} + \frac{ 4 M Q^2}{r^{3}} + \frac{1}{r^2} \left( - \frac{M^2}{4} + 8 m^2 Q^2 - 8 Q^2 \Lambda  - 2 m^2 M r \right) + ( 8 m^2 \Lambda + 4 \Lambda^2) 
\end{equation}
$V(r)$ is plotted in Fig.1.
\begin{center}

\scalebox{0.9} {\includegraphics{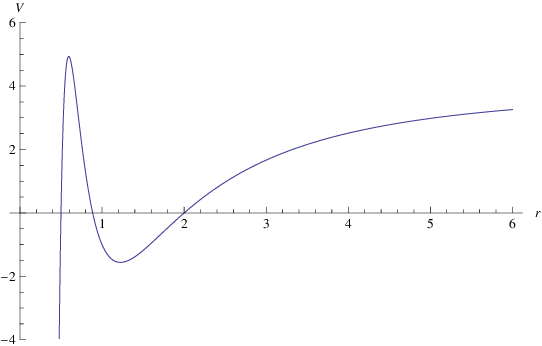}}

\vspace{0.3cm}

{Figure 1. The behavior of $V(r)$ with $r$ for $\Lambda=1$ $M=10$, $Q=1$ and $m=0$}

\vspace{0.3 cm}

\end{center}
The ``tortoise'' coordinate $r_*$ is given by,
\begin{equation}
r_* = \frac{ 1}{ 4 \Lambda (r_+ - r_-)} \left( r_+ ln( (r - r_+) - r_- ln( r - r_-) \right)
\end{equation}
Note that when $r \rightarrow r_+$ , $r_* \rightarrow - \infty$ and for $r \rightarrow \infty$, $ r_* \rightarrow \infty $.

In order to solve the wave equation exactly, one can redefine the 
$r$ coordinate of the eq.(15) with a new variable $z$ given by,
\begin{equation}
 z = \left( \frac{ r - r_+}{ r - r_-} \right)
\end{equation}
Then eq.(15) becomes,
\begin{equation}
z(1-z) \frac{d^2 R}{dz^2} + (1-z) \frac{d R}{dz} + P(z) R =0
\end{equation}
Here,
\begin{equation}
P(z) = \frac{A}{z} + \frac{B}{-1+z} + C
\end{equation}
where,
\begin{equation}
A=  \frac{\omega^2 r_+^2}{ 16 (r_+- r_-)^2 \Lambda^2}; \hspace{1.0cm} B = \frac{ 8m^2 \Lambda - \omega^2} { 16 \Lambda^2}; \hspace{1.0cm} C = - \frac{r_-^2 \omega^2}{16 (r_+- r_-)^2 \Lambda^2}
\end{equation}

%%%%%%%%%%%%%%%%%%%%%%%%

\section{Uncharged dilaton black hole}

First,  we will focus on the uncharged black string solution with $Q=0$ in the metric. Note that this solution has a single horizon at $r_h= \frac{M}{4 \Lambda}$. Here, we will compute the reflection coefficients, grey body factors and decay rates for the uncharged black hole. A matching procedure is used in order to obtain the amplitudes of the in going and the out going waves at infinity. Hence the space is divided into two regions: near horizon ( $r \sim r_h$) and far region ( $r \rightarrow \infty $).

For the uncharged black hole, $P(z)$ given in eq.(22) is,
$$P(z) = \frac{A}{z} + \frac{B}{-1+z}$$
\begin{equation}
A =  \frac{\omega^2}{ 16 \Lambda^2}; \hspace{1.0cm}
B = \frac{ 8m^2 \Lambda - \omega^2} { 16 \Lambda^2}
\end{equation}
Now, with the  definition of  $R$ as,
\begin{equation}
R(z) = z^{\alpha} (1-z)^{\beta} F(z)
\end{equation}
the radial equation given in eq.(22) becomes,
\begin{equation}
z(1-z) \frac{d^2 F}{dz^2} + \left(1 + 2 \alpha - (1+ 2 \alpha + 2\beta )z \right) \frac{d F}{dz} + \left(\frac{\bar{A}}{z} + \frac{\bar{B}}{-1+z} + \bar{C}     \right) F =0
\end{equation}
where,
$$\bar{A} = A + \alpha^2$$
$$ \bar{B} = B + \beta - \beta^2$$
\begin{equation}
\bar{C} = -(\alpha+ \beta)^2
\end{equation}
The above equation resembles the hypergeometric differential equation  which is of the form \cite{math},
\begin{equation}
z(1-z) \frac{d^2 F}{dz^2} + (c  - (1+a + b )z) \frac{d F}{dz} -ab  F =0
\end{equation}
By comparing the coefficients of eq.(27) and eq.(29), one can obtain the following identities,
\begin{equation}
c = 1+ 2 \alpha
\end{equation}
\begin{equation}
a+b = 2 \alpha + 2 \beta
\end{equation}
\begin{equation}
\bar{A}= A + \alpha^2 =0; \Rightarrow \alpha= \frac{ i \omega}{ 4 \Lambda}
\end{equation}
\begin{equation}
\bar{B} = B + \beta - \beta^2=0; \Rightarrow 
\beta = \frac{1 + i \sqrt{ ( \frac{ \omega^2 - 8 m^2 \Lambda}{4 \Lambda^2} ) -1 }}{2}
\end{equation}
Note that in this paper we will only focus on the perturbations with large frequencies. Hence we will assume that 
\begin{equation}
\left( \frac{ \omega^2 - 8 m^2 \Lambda}{4 \Lambda^2} \right)  \geq 1
\end{equation}
Therefore, $\beta$ will be considered imaginary for the rest of the paper. Also, 
\begin{equation}
 ab = -\bar{C} = (\alpha + \beta)^2  
\end{equation}
From eq.(31) and eq.(35),
\begin{equation}
a= b= \alpha + \beta
\end{equation}
The solution to the hypergeometric function $F(z)$ is given by \cite{math},
\begin{equation}
F(a,b;c;z) = \frac{\Gamma(c)} {\Gamma(a) \Gamma(b)} \Sigma \frac{ \Gamma(a+n) \Gamma( b+n)}{ \Gamma(c+n)}  \frac{z^n}{n!}
\end{equation}
with a radius of convergence being the unit circle $|z| =1$.

\subsection{ Solution at the near-horizon region}

First, the solution of the wave equation closer to the horizon is analyzed. For the uncharged black hole,
\begin{equation}
z = 1 - \frac{r_h}{r}
\end{equation}
and the ``tortoise'' coordinate for uncharged black hole is
\begin{equation}
r_* = \frac{1}{ 4 \Lambda} ln( r - r_h)
\end{equation}
As the radial coordinate $r \rightarrow r_h$, $z \rightarrow 0$. In the neighborhood of $z=0$, the hypergeometric function has two linearly independent solutions givern by \cite{math}
\begin{equation}
F(a,b;c;z) \hspace{1.0cm} and \hspace{1.0cm} (1-z)^{(1-c)} F(a-c+1,b-c+1;2-c;z)
\end{equation}
Substituting the  values of $a,b,c$ in terms of $\alpha$, $\beta$, the general solution for $R(z)$ can be written as,
\begin{equation}
R(z) = C_1 z^{\alpha} (1-z)^{\beta} F(\alpha + \beta, \alpha+\beta, 1+ 2 \alpha, z) +
C_2 z^{-\alpha}(1-z)^{\beta} F( -\alpha + \beta,-\alpha+\beta,1-2 \alpha, z)
\end{equation}
Here, $C_1$ and $C_2$ are constants to be determined. Since $z \rightarrow 0$ closer to the horizon, the above solution approaches,
\begin{equation}
R(z  \rightarrow 0)= C_1 z^{\alpha}  + C_2 z^{- \alpha}
\end{equation}
Since closer to the horizon, $r \rightarrow r_h$, $z$ can be approximated with
\begin{equation}
z \approx  \frac{ r - r_h}{ r_h}
\end{equation}
Hence eq.(42) can be re-written in terms of $r_*$ as,
\begin{equation}
R(r \rightarrow r_h) =  C_1 \left(\frac{ 1} { r_h} \right)^{\alpha} e ^{ i  \omega r_*}  + C_2 \left(\frac{1}{ r_h} \right)^{-\alpha} e^{ -i  \omega r_*}
\end{equation}
The first and the second term corresponds to the ingoing and the outgoing wave respectively. Now, one can impose the condition that the wave is purely ingoing at the horizon. Hence we pick $C_1 \neq 0$ and $C_2=0$. Therefore the solution closer to the horizon is,
\begin{equation}
R(z \rightarrow 0) =  C_1 z^{\alpha} (1-z)^{\beta} F(\alpha + \beta, \alpha+\beta, 1+ 2 \alpha, z) 
\end{equation}

\subsection{Solution at asymptotic region}

When $r \rightarrow \infty$, the wave equation in eq.(15) approximates to,
\begin{equation}
\frac{d}{dr} \left( 4 \Lambda r^2  \frac{dR(r)}{dr} \right) + 2r^2 \left( \frac{\omega^2}{ 8 \Lambda r^2}   - \frac{m^2}{r^2}  \right)  R(r) =0
\end{equation}
When expanded, the equation simplifies to the Euler's equation given by,
\begin{equation}
r^2 R'' + 2 r R' + p R =0
\end{equation}
Where, $ p = -B$ given in eq.(24). The solution to the Euler equation is given by,
\begin{equation}
R(r) = D_1 \left( \frac{r_h}{r} \right)^{a_1} + D_2 \left( \frac{r_h}{r} \right)^{a_2}
\end{equation}
with,
\begin{equation}
a_1=   \frac{ 1 + \sqrt{ 1 + 4 B} }{2} =  \beta; \hspace{1.0 cm}
a_2 =  \frac{ 1 - \sqrt{ 1 + 4 B} }{2} = (1- \beta)
\end{equation}

\subsection{Matching the solutions at the near horizon and the asymptotic region}

In this section we  match the asymptotic solution given in eq.(48) to the large $r$ limit (or the $z \rightarrow 1$ ) of the near-horizon solution given in 
eq.(45).  To obtain the $z \rightarrow 1$ behavior of eq. (45), one can perform a well known transformation on hypergeometric function given as follows \cite{math}
$$
F(a,b,c,z) = \frac{ \Gamma(c) \Gamma(c-a-b)}{\Gamma(c-a) \Gamma(c-b)} F(a,b;a+b-c+1;1-z) 
$$
\begin{equation}
 +(1-z)^{c-a-b}\frac{ \Gamma(c) \Gamma(a+b-c)}{\Gamma(a) \Gamma(b)} F(c-a,c-b;c-a-b+1;1-z)
\end{equation}
Applying this transformation to eq.(45) and substituting for the values of $a,b,c$, one can obtain the solution to the wave equation in the asymptotic region as follows;
$$
R(z) = C_1 z^{\alpha} (1-z)^{\beta} \frac{ \Gamma(1 + 2 \alpha) \Gamma(1 - 2 \beta)}{\Gamma(1 + \alpha - \beta)^2} F( \alpha + \beta, \alpha + \beta ; 2 \beta - 1;1-z) $$
\begin{equation}
+ C_1  z^{\alpha} (1-z)^{1 - \beta} \frac{ \Gamma( 1 + 2 \alpha ) \Gamma( -1 + 2 \beta )}{\Gamma( \alpha + \beta )^2 } F( 1 + \alpha - \beta , 1 + \alpha - \beta ; 2 - 2 \beta;1-z)
\end{equation}
Now we can take the limit of  $R(z)$ as $ z \rightarrow 1$ which  lead to,
\begin{equation}
R(z \rightarrow 1) =  C_1  (1-z)^{\beta} \frac{ \Gamma(1 + 2 \alpha) \Gamma(1 - 2 \beta)}{\Gamma(1 + \alpha - \beta)^2} 
+ C_1  (1-z)^{1 - \beta} \frac{ \Gamma( 1 + 2 \alpha ) \Gamma( -1 + 2 \beta )}{\Gamma( \alpha + \beta )^2 } 
\end{equation}
Since,
\begin{equation}
1 - z = \frac{r_h}{r}
\end{equation}
$R(r)$ for $ r \rightarrow \infty$ (or $ z \rightarrow 1$) can be written as,
\begin{equation}
R(r \rightarrow \infty ) = C_1  \left(\frac{r_h}{r}\right)^{\beta} \frac{ \Gamma(1 + 2 \alpha) \Gamma(1 - 2 \beta)}{\Gamma(1 + \alpha - \beta)^2} 
+ C_1  \left(\frac{r_h}{r}\right)^{1 - \beta} \frac{ \Gamma( 1 + 2 \alpha ) \Gamma( -1 + 2 \beta )}{\Gamma( \alpha + \beta )^2 } 
\end{equation}
By comparing eq. (48) and eq. (54), the coefficients $D_1$ and $D_2$ can be written as,
\begin{equation}
D_1 = C_1 \frac{ \Gamma(1 + 2 \alpha) \Gamma(1 - 2 \beta)}{\Gamma(1 + \alpha - \beta)^2} 
\end{equation}
\begin{equation}
D_2 = C_1  \frac{ \Gamma( 1 + 2 \alpha ) \Gamma( -1 + 2 \beta )}{\Gamma( \alpha + \beta )^2 } 
\end{equation}
To determine the ingoing and outgoing wave at large $r$ ( or large $r_*$), we will rewrite the $R(r)$ in terms of $r_*$  as follows:
\begin{equation}
R(r \rightarrow \infty  ) = D_1 r_h^{\beta}  e ^{ -i  \omega r_* \sqrt{1 - \frac{ 4 \Lambda^2}{ \omega^2} ( \frac{2 m^2}{\Lambda} +1) } - 2 \Lambda r_*}  + D_2 r_h^{(1-\beta)} e ^{ i  \omega r_* \sqrt{1 - \frac{ 4 \Lambda^2}{ \omega^2} ( \frac{2 m^2}{\Lambda} +1) } - 2 \Lambda r_*}  
\end{equation}
Note that $r = r_h + e^{ 4 \Lambda r_*}$ which can be approximated as $r \sim e^{ 4 \Lambda r_*}$ at large $r$. From the above it is clear that $D_1$ and $D_2$ represents the outgoing and ingoing waves respectively.

\subsection{Reflection Coefficient}

The reflection coefficient $\Re$ can be defined as,
\begin{equation}
\Re =  \left|\frac{D_1}{D_2} \right|^2 = \frac{Cosh^2( 
\frac{ \pi \omega}{ 4 \Lambda} - \frac{\pi }{2} \sqrt{ 
\frac{ (\omega^2 - 8 m^2 \Lambda)}{4 \Lambda^2} - 1 } ) } { Cosh^2( 
\frac{ \pi \omega}{ 4 \Lambda} + \frac{\pi }{2} \sqrt{ 
\frac{ (\omega^2 - 8 m^2 \Lambda)}{4 \Lambda^2} - 1 } )}
\end{equation}
Note that we have used the identities for $\Gamma$ functions given by,
\begin{equation}
|\Gamma( i y )|^2 =  \frac{ \pi } { y Sinh ( \pi y ) }; \hspace{0.3cm} 
|\Gamma( \frac{1}{2} + i y )|^2 = \frac{ \pi} { Cosh ( \pi y ) }; \hspace{0.3 cm} 
|\Gamma( 1 + i y )|^2 = \frac{ \pi y } { Sinh ( \pi y )}
\end{equation}

In the limit of large energies or large $\omega$, the reflection coefficient approaches zero as shown in Fig.2. This is in agreement with the fact that at high energies the black hole absorbs more than at smaller energies.

\begin{center}

\scalebox{0.9} {\includegraphics{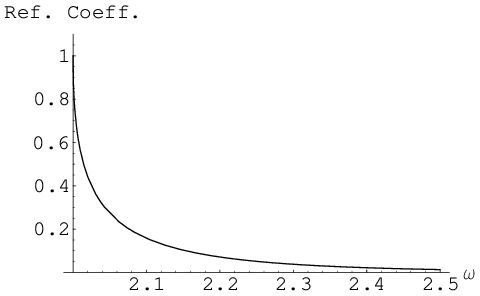}}
\vspace{0.3cm}

{Figure 2. The behavior of $\Re$ with $\omega$ for $\Lambda=1$, $m=0$}
\vspace{0.3 cm}

\end{center}

\subsection{Flux calculation}

The absorption cross section also can be obtained by computing the flux at the horizon and infinity. First we will define the conserved flux for the wave function as,
\begin{equation}
F = \frac{2 \pi} {i} \left ( R^{*} f(r) \frac{dR(r)}{dr} - R(r) f(r) \frac{ dR^*}{dr} \right)
\end{equation}
The function $f(r)= 8 \Lambda(r - r_h) r$. Using the function $R(r) = C_1 z^{\alpha}$ at the horizon, the ingoing flux at the horizon is,
\begin{equation}
F(r \rightarrow r_h) = 8 \pi |C_1|^2 \omega r_h
\end{equation}
Using the eq.(57) for the function $R(r)$ at infinity, the incoming flux at infinity is given by,
\begin{equation}
F(r \rightarrow \infty) = 16 \pi |D_2|^2 \Lambda r_h \sqrt{\frac{ (\omega^2 - 8 m^2 \Lambda)}{4 \Lambda^2} - 1 }
\end{equation}
The partial wave absorption is,
\begin{equation}
\sigma = \frac{ F(r \rightarrow r_h)}{F(r \rightarrow \infty)} = \frac{ \omega }{ 2 \Lambda  \sqrt{\frac{ (\omega^2 - 8 m^2 \Lambda)}{4 \Lambda^2} - 1 } } \left| \frac{\Gamma( \alpha + \beta)^2 }{ \Gamma( 1 + 2 \alpha) \Gamma( -1 + 2 \beta)} \right|^2
\end{equation}
By substituting the values of $\alpha$ and $ \beta$ the above expression simplifies to,
\begin{equation}
\sigma= \frac{ Sinh \left( \pi \sqrt{ \frac{ (\omega^2 - 8 m^2 \Lambda)}{4 \Lambda^2} - 1  } \right)  Sinh \left(  \frac{ \pi \omega}{ 2 \Lambda}  \right)}
{Cosh^2 \left( 
\frac{ \pi \omega}{ 4 \Lambda} + \frac{\pi }{2} \sqrt{ 
\frac{ (\omega^2 - 8 m^2 \Lambda)}{4 \Lambda^2} - 1 }  \right)}
\end{equation}
Note that we will take the absolute value of the  expression $\left(\frac{ (\omega^2 - 8 m^2 \Lambda)}{4 \Lambda^2} - 1  \right) $ when computing the cross sections. Otherwise $\beta$ would be real and the flux at infinity would be zero. 

\subsection{Absorption cross sections and decay rates}

The absorption cross section( or the grey body factor)  in three dimensions is $ \sigma_{abs} = \sigma/\omega$. This was clearly defined in Das et. al. \cite{das2}. Hence,
\begin{equation}
\sigma_{abs} = \frac{ \left( e^{ \frac{ \pi \omega}{\Lambda}} -1  \right) \left( e^{ 2\pi \sqrt{ 
\frac{ (\omega^2 - 8 m^2 \Lambda)}{4 \Lambda^2} - 1 } } - 1 \right)}{\omega  \left(e^{ \frac{\pi \omega}{2 \Lambda} + \pi \sqrt{ 
\frac{ (\omega^2 - 8 m^2 \Lambda)}{4 \Lambda^2} - 1 }} + 1 \right)^2}
\end{equation}
We will plot $\sigma_{abs}$ in Fig.3.

\begin{center}

\scalebox{0.9} {\includegraphics{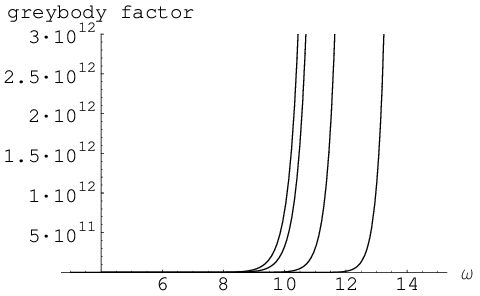}}
\vspace{0.3cm}

{Figure 3. The behavior of $\sigma_{abs}$ with $\omega$ for $\Lambda=1$}
\vspace{0.3 cm}

\end{center}
The absorption cross section diverge with increasing $\omega$. The three graphs are for $m=0,1,2$. Here, $\sigma_{abs}$ graph shift towards right with increasing $m$.The graphs are drawn in such a way that $\omega^2 \geq 8 m^2 \Lambda + 4 \Lambda^2$ 

For  static and spherically symmetric black hole in any dimensions,
the grey body factor for a minimally coupled scalar field when $\omega \rightarrow 0$ was shown to be  equal to the horizon area by Das et. al. \cite{ das2}. Since we have restricted $\omega$ with eq.(34) to be positive, one can not take such a limit to prove this for the absorption cross sections considered in this paper. How ever, if one wants to take  the $\omega \rightarrow 0$ limit then one has to remember that $\beta$ in eq. (33) will become real. Hence, flux of the asymptotic solution to the wave function will be zero. One remedy for this problem is to follow the procedure followed by Birmingham et.al. \cite{bir1} in computing the low frequency absorption cross section of the BTZ black hole for scalar perturbations. There, they combined the two independent solutions of the wave equation to give a complex function with positive and negative flux at infinity. We will leave this for further study. Now,  the decay rate for this black hole is,
\begin{equation}
\Gamma_{decay} = \frac{\sigma_{abs}} { e^{\omega/T} -1}
\end{equation}
Here the temperature $T = \frac{\pi}{\Lambda}$ for the uncharged black hole. Hence,
\begin{equation}
\Gamma_{decay} = \frac{  \left( e^{ 2\pi \sqrt{ 
\frac{ (\omega^2 - 8 m^2 \Lambda)}{4 \Lambda^2} - 1 }}  - 1 \right)}{\omega  \left(e^{ \frac{\pi \omega}{2 \Lambda} + \pi \sqrt{ 
\frac{ (\omega^2 - 8 m^2 \Lambda)}{4 \Lambda^2} - 1 } } + 1 \right)^2}
\end{equation}

\begin{center}

\scalebox{0.9} {\includegraphics{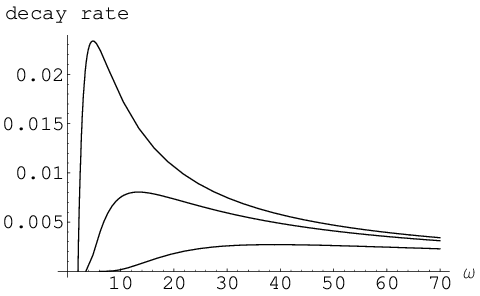}}

\vspace{0.3cm}

{Figure 4. The behavior of $\Gamma$ with $\omega$ for $\Lambda=1$}

\vspace{0.3 cm}

\end{center}

We plot $\Gamma$ in Fig.4. The three graphs are for $m=0,1,2$. $\Gamma$ becomes smaller for increasing $m$.

\subsection{Quasi normal modes of uncharged black hole}

Quasi normal modes of a classical perturbation of black hole space-times are defined as the solutions to the related wave equations with purely ingoing waves at the horizon. In addition, one has to impose boundary conditions on the solutions at the asymptotic regions as well. In asymptotically flat space-times, the second boundary condition is the solution to be purely outgoing  at spatial infinity. For asymptotically AdS space times, various boundary conditions have been chosen in the current literature. One of the choices is the  Dirichelet boundary condition. Another condition for AdS spaces is to  make the energy-momentum flux density to vanish asymptotically \cite{bir2}. Here, we will show that both condition leads to the same values for quasi normal modes.
First, by the Dirichelet condition, the amplitude of the ingoing wave has to be zero. This leads to, $D_2=0$. Second, the vanishing of incoming flux at infinity $ F_{\infty}$ computed in eq.(62) leads to $|D_2|^2=0$. Hence, the quasi normal modes are obtained for,

\begin{equation}
D_2 = \frac{ \Gamma( 1 + 2 \alpha ) \Gamma( -1 + 2 \beta )}{\Gamma( \alpha + \beta )^2 } =0
\end{equation}
Note that the  Gamma function $\Gamma(x)$ has poles at $x=-n$ for $ n=0,1,2..$. Hence  the  function $D_2$ will vanish at the poles of $\Gamma( \alpha + \beta)$ given by,
\begin{equation}
\alpha + \beta  = -n
\end{equation}
By combining eq.(32), eq.(33) and eq.(69)
leads to the quasi normal frequency $\omega$ as,
\begin{equation}
\omega =  \frac{-2 i}{ 2n +1} \left( 2 \Lambda n (1+n) - m^2 \right)
\end{equation}
Note that the QNM's of this black holes have the interesting property being ``pure-imaginary''. Another interesting point to note is that the Hawking temperature of the uncharged black hole, $T_H =\Lambda/\pi$  is independent of the black hole mass. $|\omega|$ is also independent of the black hole mass. Furthermore, $|\omega|$ $\sim \Lambda$ for small $m$ values leading to a linear relation between $\omega$ and $T_H$. The quasi normal modes are discussed in detail in \cite{fer}.

\section{ Charged dilaton black hole}

In this section we will focus on the charged black hole with two horizons at $r=r_-, r_+$. The radial equation in eq.(15) simplifies into 
\begin{equation}
z(1-z) \frac{d^2 F}{dz^2} + \left(1 + 2 \alpha - (1+ 2 \alpha + 2\beta )z \right) \frac{d F}{dz} + \left(\frac{\bar{A}}{z} + \frac{\bar{B}}{-1+z} + \bar{C}\right) F =0
\end{equation}
similar to the one in the uncharged case. Here,
$$\bar{A} = A + \alpha^2$$
$$ \bar{B} = B + \beta - \beta^2$$
\begin{equation}
\bar{C} = C - (\alpha+ \beta)^2
\end{equation}
The above equation resembles the hypergeometric differential equation of the format given in eq.(29) in the previous section. By comparing the coefficients of eq.(29) and eq. (71), one can obtain the following identities,
\begin{equation}
c = 1+ 2 \alpha
\end{equation}
\begin{equation}
a+b = 2 \alpha + 2 \beta
\end{equation}
\begin{equation}
\bar{A}=A + \alpha^2 =0; \Rightarrow \alpha= \frac{  i \omega r_+}{ 4 \Lambda ( r_+ - r_-)}
\end{equation}
\begin{equation}
\bar{B} = B - \beta + \beta^2=0; \Rightarrow \beta = \frac{1 + i \sqrt{ ( \frac{ \omega^2 - 8 m^2 \Lambda}{4 \Lambda} -1 }) }{2}
\end{equation}
\begin{equation}
 ab = -\bar{C} = (\alpha + \beta)^2 - C 
\end{equation}
From eq.(74) and eq.(77),
$$a= \alpha + \beta + \gamma$$
\begin{equation}
b= \alpha + \beta -\gamma
\end{equation}
Here, 
\begin{equation}
\gamma= i\sqrt{|C|} = \frac{  i \omega r_-}{ 4 \Lambda ( r_+ - r_-)}
\end{equation}

\subsection{ Solution at the near-horizon region}

First, the solution of the wave equation closer to the horizon is analyzed. For the charged black hole,
\begin{equation}
z = \frac{(r- r_+)}{ ( r - r_-)}
\end{equation} 
and as the radial coordinate $r$ approaches the horizon, $z$ approaches $0$. In the neighborhood of $z=0$, the hypergeometric function has two linearly independent solutions given by \cite{math}
\begin{equation}
F(a,b;c;z) \hspace{1.0cm} and \hspace{1.0cm}(1-z)^{(1-c)} F(a-c+1,b-c+1;2-c;z)
\end{equation}
Substituting the  values of $a,b,c$ in terms of $\alpha$, $\beta$, and $\gamma$, the general solution for $R(z)$ can be written as,
$$
R(z) = C_1 z^{\alpha} (1-z)^{\beta} F(\alpha + \beta + \gamma, \alpha+\beta - \gamma, 1+ 2 \alpha, z) 
$$
\begin{equation}
+ C_2 z^{-\alpha}(1-z)^{\beta} F( -\alpha + \beta + \gamma,-\alpha+\beta - \gamma,1-2 \alpha, z)
\end{equation}
Here, $C_1$ and $C_2$ are constants to be determined. Since  closer to the horizon $z \rightarrow 0$, the above solution approaches,
\begin{equation}
R(z  \rightarrow 0) = C_1 z^{\alpha}  + C_2 z^{-\alpha}
\end{equation}
Closer to the horizon, $r \rightarrow r_+$. Hence, $z$ can be approximated with
\begin{equation}
z= \frac{ r - r_+}{r_+ -  r_-}
\end{equation}
The ``tortoise'' coordinate for the charged black hole is given in eq.(20).
Near the horizon $r \rightarrow r_+$, the ``tortoise'' coordinate can be approximated to be 
\begin{equation}
r_* \approx  \frac{r_+}{4 \Lambda ( r_+ - r_-)} ln( r  - r_+)
\end{equation}
Hence eq.(83) can be re-written in terms of $r_*$ as,
\begin{equation}
R(r \rightarrow r_+) =  C_1 \left(\frac{ 1} { (r_+ - r_-)} \right)^{\alpha} e ^{ i  \omega r_*}  + C_2 \left(\frac{1}{ (r_+ - r_-)} \right)^{ - \alpha} e^{ -i  \omega r_*}
\end{equation}
The first and the second term corresponds to the ingoing and the outgoing wave respectively. Now, one can impose the condition  that  the wave is purely ingoing at the horizon. Hence we pick $C_1 \neq 0$ and $C_2=0$. Therefore the solution closer to the horizon is,
\begin{equation}
R(z \rightarrow 0) ) =  C_1 z^{\alpha} (1-z)^{\beta} F(\alpha + \beta + \gamma, \alpha+\beta - \gamma, 1+ 2 \alpha, z) 
\end{equation}

\subsection{Solution at asymptotic region}

When $r \rightarrow \infty$, the behavior of the wave function is similar to the one for the uncharged case given in section 4. Note that  the value of $\beta$ in $R( r \rightarrow \infty)$ is the same for both the uncharged and charged case.

\subsection{Matching the solutions at the near horizon and the asymptotic region}

In this section we  match the asymptotic solution given in eq.(48) to the large $r$ limit (or the $z \rightarrow 1$ ) of the near-horizon solution given in 
eq.(87).  To obtain the $z \rightarrow 1$ behavior of eq.(87), one can follow  similar procedure as for the uncharged case given in section 4. Substituting for the values of $a,b,c$ for the charged black hole, one can obtain the solution to the wave equation in the asymptotic region as follows;
$$
R(z) = C_1 z^{\alpha} (1-z)^{\beta} \frac{ \Gamma(1 + 2 \alpha) \Gamma(1 - 2 \beta)}{\Gamma(1 + \alpha - \beta - \gamma) \Gamma( 1 + \alpha - \beta + \gamma)} F( \alpha + \beta + \gamma, \alpha + \beta -\gamma; 2 \beta ;1-z) $$
\begin{equation}
+ C_1  z^{\alpha} (1-z)^{1 - \beta} \frac{ \Gamma( 1 + 2 \alpha ) \Gamma( -1 + 2 \beta )}{\Gamma( \alpha + \beta + \gamma) \Gamma( \alpha + \beta - \gamma)} F( 1 + \alpha - \beta -\gamma, 1 + \alpha - \beta + \gamma ; 2 - 2 \beta;1-z)
\end{equation}
Now we can take the limit of the above $R(z)$ as $ z \rightarrow 1$ which will lead to,
$$
R(z \rightarrow 1) = C_1  (1-z)^{\beta} \frac{ \Gamma(1 + 2 \alpha) \Gamma(1 - 2 \beta)}{\Gamma(1 + \alpha - \beta - \gamma) \Gamma( 1 + \alpha - \beta + \gamma)} 
$$
\begin{equation}
+ C_1 (1-z)^{1 - \beta} \frac{ \Gamma( 1 + 2 \alpha ) \Gamma( -1 + 2 \beta )}{\Gamma( \alpha + \beta + \gamma) \Gamma( \alpha + \beta - \gamma) } \end{equation}
Since,
\begin{equation}
1 - z = \frac{r_+ - r_-}{r- r_-}
\end{equation}
$R(r)$ for large $r$ can be written as,
$$
R(r \rightarrow \infty) = C_1  \left(\frac{r_+ - r_-}{r - r_-}\right)^{\beta} \frac{ \Gamma(1 + 2 \alpha) \Gamma(1 - 2 \beta)}{\Gamma(1 + \alpha - \beta - \gamma) \Gamma( 1 + \alpha - \beta + \gamma}
$$
\begin{equation} 
+ C_1  \left(\frac{r_+ - r_-}{r - r_-}\right)^{1 - \beta} \frac{ \Gamma( 1 + 2 \alpha ) \Gamma( -1 + 2 \beta )}{\Gamma( \alpha + \beta + \gamma) \Gamma( \alpha + \beta - \gamma) } 
\end{equation}
By comparing eq.(48) and eq.(91), the coefficients $D_1$ and $D_2$ can be written as,
\begin{equation}
D_1 = C_1 \frac{ \Gamma(1 + 2 \alpha) \Gamma(1 - 2 \beta)}{\Gamma(1 + \alpha - \beta + \gamma) \Gamma( 1 + \alpha - \beta - \gamma) } 
\end{equation}
\begin{equation}
D_2 = C_1  \frac{ \Gamma( 1 + 2 \alpha ) \Gamma( -1 + 2 \beta )}{\Gamma( \alpha + \beta + \gamma) \Gamma( \alpha + \beta - \gamma )} 
\end{equation}
To determine the ingoing and outgoing wave at large $r$ ( or large $r_*$), we will rewrite the $R(r)$ in terms of ($r_*$ $ \omega$ ) as follows:
\begin{equation}
R(r  \rightarrow \infty ) \rightarrow D_1  e ^{ -i  \omega r_* \sqrt{1 - \frac{ 4 \Lambda^2}{ \omega^2} ( \frac{2 m^2}{\Lambda} +1) } - 2 \Lambda r_*} (r_+ - r_-)^{\beta} + D_2 e ^{ i  \omega r_* \sqrt{1 - \frac{ 4 \Lambda^2}{ \omega^2} ( \frac{2 m^2}{\Lambda} +1) } - 2 \Lambda r_*} (r_+ - r_-)^{(1-\beta)} 
\end{equation}
Note that we have taken $r \approx e^{ 4 \Lambda r_*}$ at large $r$. From the above it is clear that $D_1$ and $D_2$ represents the outgoing and ingoing wave respectively.

\subsection{Reflection Coefficient}

The reflection coefficient $\Re$ for the charged black hole is,

\begin{equation}
\Re= \left|\frac{ D_2}{D_1} \right|^2 =
\frac{  Cosh \left( 
\frac{  \pi \omega}{ 4 \Lambda} - \frac{\pi }{2} \sqrt{ 
\frac{ (\omega^2 - 8 m^2 \Lambda)}{4 \Lambda^2} - 1 } \right)  Cosh \left( 
\frac{ \pi \omega (r_+ + r_-)}{ 4 \Lambda ( r_+ - r_-)} - \frac{\pi }{2} \sqrt{ 
\frac{ (\omega^2 - 8 m^2 \Lambda)}{4 \Lambda^2} - 1 } \right) }
{  Cosh \left( 
\frac{ \pi \omega}{ 4 \Lambda} + \frac{1}{2} \sqrt{ 
\frac{ (\omega^2 - 8 m^2 \Lambda)}{4 \Lambda^2} - 1 } \right)  Cosh \left( 
\frac{ \pi \omega (r_+ + r_-)}{ 4 \Lambda ( r_+ - r_-)} + \frac{\pi }{2} \sqrt{ 
\frac{ (\omega^2 - 8 m^2 \Lambda)}{4 \Lambda^2} - 1 } \right) }
\end{equation}
In the limit of large energies or large $\omega$, the reflection coefficient approaches zero as shown in Fig.5. This is in agreement with the fact that at high energies the black hole absorbs more than at smaller energies.

\begin{center}

\scalebox{0.9} {\includegraphics{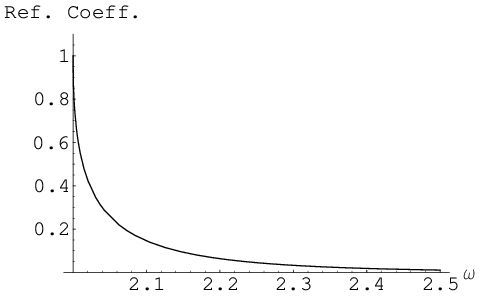}}
\vspace{0.3cm}

{Figure 5. The behavior of $\Re$ with $\omega$ for $\Lambda=1$, $m=0$, $r_- = 1$, $r_+ =3$}
\vspace{0.3 cm}

\end{center}

\subsection{Flux calculation}

The absorption cross section  can be obtained by computing the flux at the horizon and infinity. The procedure is similar to what was described for the uncharged black hole in section 4. The ingoing flux at the horizon is,
\begin{equation}
F( r \rightarrow  r_+) = 8 \pi |C_1|^2 \omega ( r_+ - r_-)
\end{equation}
Using the eq.(94) for the function $R(r)$ at infinity, the incoming flux at infinity is given by,
\begin{equation}
F( r \rightarrow \infty) = 16  \pi |D_2|^2 \Lambda \sqrt{\frac{ (\omega^2 - 8 m^2 \Lambda)}{4 \Lambda^2} - 1 }
\end{equation}
The partial wave absorption is,
\begin{equation}
\sigma = \frac{ F( r \rightarrow  r_+)}{F( r \rightarrow  \infty)} = \frac{ \omega }{ 2   \Lambda \sqrt{\frac{ (\omega^2 - 8 m^2 \Lambda)}{4 \Lambda^2} - 1 } } \left| \frac{\Gamma( \alpha + \beta + \gamma) \Gamma( \alpha + \beta - \gamma)}{ \Gamma( 1 + 2 \alpha) \Gamma( -1 + 2 \beta) } \right|^2
\end{equation}
By substituting the values of $\alpha$, $ \beta$ and $\gamma$  the above expression simplifies to,

\begin{equation}
\sigma= \frac{ Sinh \left( \pi \sqrt{ \frac{ (\omega^2 - 8 m^2 \Lambda)}{4 \Lambda^2} - 1  } \right)  Sinh \left( \frac{ \pi \omega r_+}{ 2 \Lambda (r_+ - r_-)} \right)}
{  Cosh \left(
\frac{ \pi \omega}{ 4 \Lambda} + \frac{\pi }{2} \sqrt{ 
\frac{ (\omega^2 - 8 m^2 \Lambda)}{4 \Lambda^2} - 1 }  \right) Cosh \left( 
\frac{ \pi \omega (r_+ + r_-)}{ 4 \Lambda ( r_+ - r_-)} + \frac{\pi }{2} \sqrt{ 
\frac{ (\omega^2 - 8 m^2 \Lambda)}{4 \Lambda^2} - 1 }  \right)}
\end{equation}
Note that we will take the absolute value of the  expression $\left( \frac{ (\omega^2 - 8 m^2 \Lambda)}{4 \Lambda^2} - 1  \right)$  when computing the cross sections. Otherwise $\beta$ would be real and the flux at infinity would be zero. 

\subsection{Absorption cross sections and decay rates}

The absorption cross section( or the grey body factor)  in three dimensions $ \sigma_{abs} = \sigma/\omega$. 

\begin{equation}
\sigma_{abs} = \frac{ \left( e^{ \frac{ \pi \omega r_+}{\Lambda ( r_+ - r_+)}} -1 \right) \left( e^{ 2\pi  \sqrt{ 
\frac{ (\omega^2 - 8 m^2 \Lambda)}{4 \Lambda^2} - 1 } } - 1 \right)}{\omega  \left(e^{ \frac{\pi \omega (r_+ + r_-)}{2 \Lambda (r_+ - r_-)} + \pi \sqrt{ 
\frac{ (\omega^2 - 8 m^2 \Lambda)}{4 \Lambda^2} - 1 }} + 1 \right) \left(e^{ \frac{\pi \omega r_+}{ \Lambda (r_+ - r_-)} + \pi \sqrt{ 
\frac{ (\omega^2 - 8 m^2 \Lambda)}{4 \Lambda^2} - 1 }} + 1 \right)  }
\end{equation}
We will plot $\sigma_{abs}$  in Fig.6.

\begin{center}

\scalebox{0.9} {\includegraphics{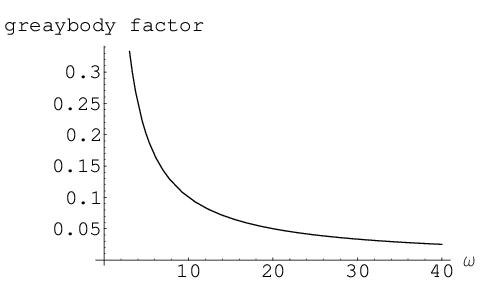}}
\vspace{0.3cm}

{Figure 6. The behavior of $\sigma_{abs}$ with $\omega$ for $\Lambda=1$}
\vspace{0.3 cm}

\end{center}
Finally the decay rate for this black hole is given by,
\begin{equation}
\Gamma = \frac{\sigma_{abs}} { e^{\omega/T} -1}
\end{equation}
The temperature for the charged black hole is 
\begin{equation}
T = \frac {\Lambda ( r_+ - r_-)} { \pi r_+}
\end{equation}
Hence $\Gamma$ simplifies to,
\begin{equation}
\Gamma = \frac{ \left( e^{ 2\pi  \sqrt{ 
\frac{ (\omega^2 - 8 m^2 \Lambda)}{4 \Lambda^2} - 1 } } - 1 \right)}{\omega  \left(e^{ \frac{\pi \omega (r_+ + r_-)}{2 \Lambda (r_+ - r_-)} + \pi \sqrt{ 
\frac{ (\omega^2 - 8 m^2 \Lambda)}{4 \Lambda^2} - 1 }} + 1 \right) \left(e^{ \frac{\pi \omega r_+}{ \Lambda (r_+ - r_-)} + \pi \sqrt{ 
\frac{ (\omega^2 - 8 m^2 \Lambda)}{4 \Lambda^2} - 1 }} + 1 \right)  }
\end{equation}
$\Gamma$ is plotted in Fig.7.

\begin{center}

\scalebox{0.9} {\includegraphics{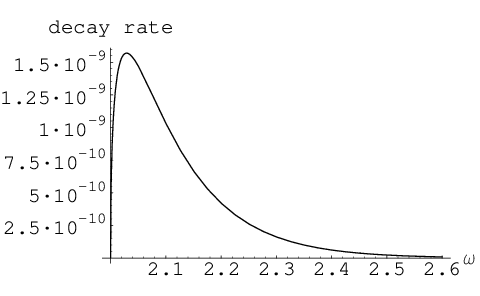}}
\vspace{0.3cm}

{Figure 7. The behavior of $\Gamma$ with $\omega$ for $\Lambda=1$}

\vspace{0.3 cm}

\end{center}

\subsection{Quasi normal modes of charged black hole}

Similar to the uncharged case, quasi normal modes are obtained by imposing the Dirichelet condition or the vanishing flux at infinity. Both leads to $D_2=0$. Observation of $D_2$ shows that it will be zero at the poles of $\Gamma(\alpha + \beta + \gamma)$ and $\Gamma( \alpha + \beta - \gamma)$ leading to,
\begin{equation}
\alpha + \beta \pm \gamma = -n
\end{equation}
Due to the possibility of $``\pm"$ in the eq.(104), there are two possibilities for $\kappa$  given in the following equations;
\begin{equation}
 \beta + \kappa \omega = -n
\end{equation}
where,
\begin{equation}
\kappa =  1,   \left( \frac{r_+ + r_-}{r_+ - r} \right)
\end{equation} 
 By eliminating $\beta$ from eq.(104) and eq.(105) leads to the quasinormal frequency $\omega$ as,
\begin{equation}
\omega=   \frac{- 2 i}{ ( \kappa^2 - 1)} \left(  \kappa \Lambda(1+2 n) \pm \sqrt{\Lambda} \sqrt{ 2m^2 	(\kappa^2 -1) + \Lambda(\kappa^2 + 4 n  (1+n))}    \right)
\end{equation}
Note that  $\omega$ for $\kappa = 1$ is same  as for the uncharged case. Hence  it  was eliminated without loss of generality.  Since $\kappa >1$, $\omega$ will always be pure imaginary. Hence for the charged black hole  the QNM's are pure imaginary similar to the uncharged one.

\section{Conclusion}
We have computed the grey body factors, reflection coefficients  and decay rates of the static dilaton black holes in 2+1 dimensions. We have also discussed the QNM's of both black holes. The special property of this computation as compared to other work in which decay rates were computed is that this is an exact calculation.

The work here motivate studies in various directions. The decay rates of the BTZ black hole was shown to match exactly with the  conformal field theoretical  description by Birmingham et. al. \cite{bir1}. It would be interesting to compute the decay rates from  the related  conformal theoretical description of this black hole and to compare with the semi classical results given in this paper. Also as mentioned in the section 4 and  5  of this paper, the computations done in this paper imposes a lower bound on $\omega$. The next step is to compute the low energy decay rates for this black hole. The steps followed by Birmingham et.al. \cite{bir1}  would lead to low energy decay rates.

As noted in section 2, the uncharged black hole and the charged black hole considered in this paper are dual to each other. Since duality plays an important role in string theory it may interesting to study the relation of the semi classical results of the charged and uncharged black holes.

Another  suggestion for further study directs towards the charged black hole obtained in Horowitz and Welch \cite{horo2}. This was obtained  by applying a duality transformation to the BTZ black hole. A natural question is how the decay rates of the  charged black hole discussed  here and the one in \cite{horo2} are related.

In this work supersymmetry has been ignored. An important issue is to find the appropriate supergravity theory in which dilaton  black hole could  be embedded.
If such a theory exists does extremality play a role in it ?. If so how does the extreme black hole decay ?. Along these lines, one may study the    decay rates of fermionic and vector  perturbations around extreme   black holes to understand any relation among them.

\end{document}